\documentclass[prl,nofootinbib,reprint]{revtex4-1}
\usepackage[dvipdfm]{graphicx}
\usepackage{amsfonts}
\usepackage{amssymb}
\usepackage{amsmath}
\usepackage{bm}
\usepackage{graphicx}
\usepackage{multirow}
\usepackage{color}
\usepackage{graphics}
\usepackage{caption}
\usepackage{epsfig}
\usepackage{dcolumn}
\usepackage{rotating}
\usepackage{color}
\usepackage{hyperref}

\newcommand{\beq}{\begin{equation}}
\newcommand{\eeq}{\end{equation}}
\def\be#1\ee{\begin{align}#1\end{align}}
\newcommand{\ov } {\over }

\begin{document}

\title{On UltraViolet effects in protected inflationary models}

\author{Diego Chialva \\}
\affiliation{
\vspace{1mm}
{\small
 Universit\'e de Mons, Service de M\'ecanique et gravitation, Place
du parc 20, 7000 Mons, Belgium}}




\begin{abstract}

Inflationary models are usually UV sensitive. Several mechanism have
been proposed to protect the necessary
features of the potential,
and most notably (softly broken) global
symmetries as shift-symmetry.
We show that,
even in presence of these protecting mechanisms, the models maintain a serious UV-dependence.
Via an improved effective theory analysis, we show how these
corrections could significantly affect
the duration of inflation,
its robustness against the choice of initial conditions
and the regimes that make
it possible.


\end{abstract}

\maketitle


Inflation is a period of accelerated expansion of the universe conjectured to
set up suitable initial conditions for the Hot Big Bang,
alleviating its fine-tuning issues \cite{Inflation, InflationReview}.
It also provides initial conditions (the field
perturbations) for structure formation.
When quantized, these structure seeds
are
natural and not fine tuned \cite{QuantumPerturbations}.

The long debate on 
fine-tuning and on the effects of higher-energy
physics
\cite{Inflation, InflationReview,InitialConditionsIssue,
  Remmen:2013eja, InhomogeneousInitialConditionsIssue} points to a UV
dependence of inflationary models, 
which are typically effective field theories.
Unfortunately, our candidate UV-complete theories
are not sufficiently developed
to give definite answers.
Effective
theory studies beyond
the lowest order
are thus necessary, notwithstanding the
limitations of a bottom-up
approach.
The onset of
inflation has been studied in various
models
\cite{InitialConditionsIssue, InhomogeneousInitialConditionsIssue},
and mechanisms have been proposed
to  make the fine-tuning of the
effective potential more natural \cite{InflationReview}.

We shall
however show that,
even in the presence of such mechanisms,
the models retain a serious
UV-dependence, whose effects have not been 
studied systematically.
Via an improved effective-theory
analysis, we shall
investigate quantitatively, and yet somewhat generally,
how it can
considerably affect dynamics, duration
and onset of inflation.

We conservatively root our work in the present
well-established theoretical
framework,
leaving for the future 
the analysis in
specific UV completions.

To be definite, we shall focus on scalar field models of inflation driven by
the potential energy, in
particular single-field ones\footnote{Field redefinitions can trade
  terms between potential and kinetic parts of the action,
  making their distinction somewhat arbitrary. However, 
  certain protecting mechanism 
  (for example shift-symmetry) allow to distinguish them 
  thanks to their transformation properties.
  \\
  \indent Note also that UV-sensitivity occurs
  in models driven by modified kinetic terms,
  as $k$-inflation \cite{ArmendarizPicon:1999rj, GhostInflation},
  and our analysis could be extended there.
  However, usually protecting mechanisms are less
  studied in those cases, and sometimes obstacles to UV completion
  appear \cite{Adams:2006sv}. We shall mention such models when
  relevant.}, which are among
the most studied and in very good
agreement with observations

\section{Inflation,
protecting mechanisms and residual UV sensitivity}

According to observations, the early universe is well described by
a flat Friedman-Robertson-Walker metric
 \beq \label{MetrAnsa}
  ds^2 = g_{\mu\nu}dx^\mu dx^\nu = -dt^2+a(t)^2d\vec x^2, \qquad
     H = { \dot a \ov a},
 \eeq
where $a$ is the scale factor, $H$ is the Hubble rate and dots
indicate time derivatives.

Accelerated expansion requires pressure $p$ and energy density $\rho$ such that $p < -\, \rho/3$. For the typical minimally
coupled, canonically normalized scalar field, with action
 \beq \label{StaAct}
  S = -\int d^4x \Bigl[{1 \ov 2}g^{\mu\nu}\nabla_\mu\nabla_\nu\phi + V(\phi)\Bigl],
 \eeq
accelerated expansion
is reliably obtained if
 \begin{gather} \label{SlowRoll}
    \!\!\!\!\rho, \text{-}p \!\sim\! V, \;\;
     \epsilon_{_V} \!\!=\!\! {M_{\text{Pl}}^2 \ov2} \Bigl(\!{\partial_\phi V \ov V}\!\Bigr)^{\!\!2} \!\ll\! 1,  \;\;
      |\eta_{_V}| \!\!=\!\! M_{\text{Pl}}^2 \biggl|{\partial_\phi^2 V \ov V}\biggr| \!\ll\! 1 ,
 \end{gather}
where $M_{Pl}=(8 \pi G)^{-1/2}$ is the reduced Planck mass.
Evidently
one needs to control quantum corrections in the potential. For
example, any
correction
$\delta V \!\!\sim\!\! V_{\text{tree}} {\phi^2 \ov M_{\text{Pl}}^2}$
would lead to a change
$ \delta\eta_{_V} \!\!=\!\! O(1)$, easily violating the bound on
$\eta_V$. This is the famous ``$\eta$-problem".

Typical protection mechanisms make corrections to the
potential
naturally small exploiting exact or
approximate symmetries to
control
their Wilson coefficients. Supersymmetry or global
symmetries are typical choices\footnote{We will mention the
  case of symmetries such as
  Galilean ones in footnote \ref{NecViolation}. They lead to Galilean genesis rather
than inflation.}.
However, supersymmetry has been shown not to
generally solve
the $\eta$-problem \cite{InflationReview, Supersymmetry}, since it is
necessarily broken at least at
the scale of inflation.

A widely studied
global symmetry 
is (softly broken)
shift-symmetry
$ 
  \phi \to \phi + constant 
$ 
\cite{ShiftSymmetry}.
Global symmetries are
possibly unnatural from the UV
perspective, as they
are expected to be broken by quantum gravity.
Some ways out are, for example, making shift-symmetry
discrete and relating it
to UV gauge symmetries, or 
realizing it
non-linearly \cite{RemediesShiftSymmetry}.
Despite these possible difficulties, 
(approximate) 
shift-symmetry is widely used in model building and thus an
interesting example.
Our results can be  
extended to other scenarios.

The approximate global symmetry
reduces the Wilson coefficients
of terms involving powers of the field,
but does not constrain terms
involving only field derivatives \cite{Chialva:2014rla}.
This remaining UV dependence is potentially serious.
Higher-derivative corrections affect
inflationary
perturbations (via their dispersion relations)\cite{PerturbationCorrections, ModifiedSoundSpeedBG,
  ModifiedSoundSpeed}, but, as we shall see,
also
the background inflation, altering its duration, dependence on initial conditions and predictions.

Many candidate UV-complete
theories possess distinctive high-derivative structures.
A full answer can be given only when the complete UV theory is known,
but an effective theory analysis is reliable when a
perturbative expansion is possible. In this sense, and to make the
work technically manageable, we begin by
considering a generic
inflaton Lagrangian, whose
dependence on derivatives is truncated
to first derivatives of the field:
 \beq \label{GenActFD}
  \mathcal{L} =  -M_f^4P(\phi, X) - V(\phi),
  \quad X = {g^{\mu\nu}\nabla_\mu\phi\nabla_\nu\phi \ov M_f^4},
 \eeq
where $M_f$ is the natural cutoff.
This truncation must be consistent,
as we shall discuss.
Moreover, the analysis should
include inhomogeneous initial conditions,
but for the moment we study homogeneous ones.
In fact, even with these limitations we shall
point out 
relevant effects.

There are two important points.
{\em First}, our classical analysis requires
$M_f < M_{\text{Pl}}$.
{\em Second}, in the absence of insights from the UV theory,
the low-energy theory only allows imposing generic initial conditions
up to the largest allowed values close to the cutoff 
$M_f$, and accounting for a soft breaking,
 \beq \label{GenInCond}
    |\partial\phi|^2 \lesssim M_f^4, \quad V < M_f^4
 \eeq
Finally, we recall that the system given by  (\ref{GenActFD})
and the Einstein-Hilbert action evaluated on 
the ansatz (\ref{MetrAnsa}) is Hamiltonian, hence Liouville's theorem applies and
no canonical standard attractor exist.
However, inflationary
attractors appear using certain coordinates \cite{Remmen:2013eja}.
This is the physical point: we describe the evolution of physically relevant variables.
In our case they will be those we use to write the effective theory:
$\phi$ and $\dot \phi$.

\section{Inflationary dynamics with the standard Lagrangian}\label{StAcModels}

The dynamical system obtained from (\ref{StaAct}) and
the Einstein-Hilbert action with
the ansatz (\ref{MetrAnsa}), is
 \beq \label{StanDynSis}
  \begin{cases}
   \dot \phi = y
    \\
   \dot y = -\sqrt{3}{\sqrt{{y^2 \ov 2}+V(\phi)} \ov M_{\text{Pl}}}y-\partial_\phi V(\phi)
  \end{cases},
 \eeq
using the Hamiltonian constraint
 $ 
  H^2 = {1 \ov 3 M_{\text{Pl}}^2}\Bigl({y^2 \ov 2}+V\Bigr)
 $. 

The system is effectively
two-dimensional, so that dynamical chaos, which would make the dependence on initial
conditions dramatic, cannot occur.
The only fixed
points of the system
(\ref{StanDynSis}) are of the form $(\phi, y) = (\phi_c, 0)$ with
$\partial_\phi V|_{\phi_c}=0$. Points at infinity do not occur because of the bounds
  (\ref{GenInCond}).
At the fixed points, the solution is flat (for  $V(\phi_c) = 0$) or de Sitter-like with
constant $H = \sqrt{3 V(\phi_c) \ov M_{\text{Pl}}}$. One can easily
see that the fixed points are
asymptotically stable if $(\partial^2_\phi V(\phi))|_{\phi_c} > 0$ and unstable if
$(\partial^2_\phi V(\phi))|_{\phi_c} < 0$.
However, these  attractors are not the physically interesting ones: a pure
de Sitter expansion would not yield the necessary density
perturbations to account for structures, nor it will have an
end\footnote{Ghost inflation \cite{GhostInflation}, which is driven by
  special kinetic terms, allows scalar perturbations in pure de Sitter,
  but it has
  tachyons in other backgrounds, and might
  not admit unitary UV completion.}.

In fact, the inflationary solutions arise in a different way.
We describe now a method to quickly find them, and determine their
attractor
behavior without obtaining the whole
phase-space diagram.

{\bf A)} We first determine
the regions $\mathcal{D}$, $\mathcal{I}$, $\mathcal{S}$
in $(\phi, y)$-space where $\dot \phi$ respectively decreases,
increases and is stationary.
For the system (\ref{StanDynSis}), they are individuated by 
${\sqrt{3} y \ov M_{\text{Pl}}}\!\sqrt{{y^2 \ov 2}\!\!+\!\!V}+\partial_{_{\!\!\phi}} \!V$
respectively positive, negative or zero.

{\bf B)} $\mathcal{S}$ is thus constituted of curves $y= y_\ast(\phi)$ satisfying
 $ 
  {\sqrt{3} y \ov M_{\text{Pl}}}\!\sqrt{{y^2 \ov 2}\!\!+\!\!V}\!+\!\partial_{_{\!\!\phi}} \!V = 0.
 $ 
They are {\em not} solutions of the system (\ref{StanDynSis}), but
integrating
$\dot \phi = y_\ast(\phi)$ and inserting it
in (\ref{StanDynSis}), we see
that the parts of $y_\ast(\phi)$
where the slow-roll condition
 \beq \label{SlRoIS}
    |y_\ast\partial_\phi y_\ast| \ll |y_\ast|\sqrt{{3 \ov M_{\text{Pl}}^2}\Bigl({y_\ast^2 \ov 2}+V(\phi)\Bigr)}
 \eeq
holds, are {\em good approximations} of solutions.
Inflationary trajectories occur in particular
when
$ 
   {y_\ast^2(\phi) \ov 2} \ll V(\phi).
$ 

{\bf C)} Given its definition, 
$\mathcal{S}$  contains the fixed points of the
system. Thus, since $\mathcal{S}$
separates regions $\mathcal{D}$ and $\mathcal{I}$, and the flow
draws from $\mathcal{D}$, $\mathcal{I}$ on $\mathcal{S}$, 
if there is only a global asymptotic attractor
then the approximate inflationary solutions in $\mathcal{S}$, see B),
are essentially global attractors.

{\bf C')} If there are instead multiple locally attractive fixed points,
the inflationary solutions
ending on them are
at least attractive for a local set of initial conditions.

\paragraph{\textbf{Duration and onset of inflation.}}

Even with attractor solutions,
the inflationary phase must still comply with
observations and provide correct initial conditions
for the Hot Big-Bang.
In particular, the inflationary attractors must be reached early enough and allow
a suitable duration of inflation \cite{Overshooting}.

This is partly model-dependent,
but one can discuss how fast solutions
where the kinetic
term is initially sizable
reach the inflationary regime.
For the standard system
(\ref{StanDynSis}) one
finds that all
solutions where
 $ 
    \dot \phi^2 \gg V(\phi), M_{\text{Pl}}|\partial_\phi V(\phi)|
 $ 
will damp exponentially their kinetic energy density \cite{InitialConditionsIssue},
approximately as
{\small
 \beq \label{ExpoDamp}
   \dot\phi \sim  \pm M_f^2 e^{- \sqrt{3 \ov 2}{|\phi-\phi_{\text{in}}| \ov M_{\text{Pl}}}}
 \eeq
}
This indicates that inflation would take place quite rapidly, as soon
as $|\Delta\phi| = |\phi-\phi_{\text{in}}| > M_{\text{Pl}}$. While
this argument is
not relevant for small-field
models (where the field excursion is sub-Planckian), it seems to make
the situation more favorable in  large-field
models.
We shall see shortly how this
changes when corrections are taken
into account.

\section{Dynamics with higher-order derivative
corrections}\label{HigDerCorrAcModels}

After using the Hamiltonian constraint $H^2 = {\rho \ov 3M_{\text{Pl}}}$
the field equations derived from the action (\ref{GenActFD}) read
 \beq \label{DcDynSis}
  \left\{\begin{aligned}
   \dot \phi & = y
    \\
   \dot y & = -\sqrt{3} c_s^2{\sqrt{2 P_{,X} y^2 + M_f^4 P + V} \ov M_{\text{Pl}}}y
     + {M_f^4 \ov 2}{\partial_\phi \rho \ov \partial_X \rho}
  \end{aligned}\right.
 \eeq
where commas indicate partial derivatives
and
 \be \label{EnDensHigDer}
  \rho & = 2 y^2 P_{,X}(\phi, X) + M_f^4 P(\phi, X) + V(\phi),
    \\
  c_s^2 & = {P_{,X} \ov P_{,X} - 2P_{,XX}y^2 M_f^{-4}}.
 \ee

The corrections could lead to new attractors, making suitable
inflationary solutions harder to approach
without fine-tuning of initial conditions.
Moreover, even in absence of new undesirable attractors,
they could affect the features of inflationary solutions, for instance
leading to too short a
period of inflation or to signatures incompatible with observations.

 A few basic and necessary physical conditions
suffice to constrain $P(\phi, X)$
and get concrete
results concerning these questions.
As the
complete theory is unknown, we can only consider very basic
conditions: 1) a stable and causal
system, 2) a well-posed Chauchy problem for the field dynamics
and 3) a self-consistent theory.


Stability and causality require at least the weakest
energy conditions: the null energy one
(NEC)\footnote{\label{NecViolation} There
  have been various attempts to construct reliable NEC-violating models
 in Galileon and ghost condensate scenarios, but
no totally healthy and under control
model has yet
appeared \cite{NECViolation}.}
\cite{Adams:2006sv, EnergyConditionInstability}.
From  the action (\ref{GenActFD}),
 \beq \label{HDEneMomTens}
 {\textstyle T_{\mu\nu}\equiv 
      2P_{,X}\nabla_\mu\phi\nabla_\nu\phi - g_{\mu\nu}(M_f^2P+V)},
 \eeq
and the NEC states that for any null vector $n^{\mu}$
 \beq \label{NECc}
  T_{\mu\nu}n^{\mu}n^{\nu} \geq 0 \Rightarrow P_{,X} \geq 0.
 \eeq
Other conditions follow from the covariant field equation
{\small \be \label{CovFieldEq}
   \!\!\!\!
  \Bigl(\!\!P_{_{,X}}g^{\mu\nu}+2 {\nabla^\mu\phi\nabla^\nu\phi \ov M_f^4} P_{_{,XX}}\!\!\Bigr)\nabla_{_\mu}\nabla_{_\nu}\phi+P_{_{,X\phi}} g^{\mu\nu}\nabla_{_\mu}\phi\nabla_{_\nu}\phi
   \nonumber \\
   \!\!\!\!- {(M_f^4 P+ V)_{,\phi} \ov 2} = 0.
 \ee}%
This would be generally degenerate if $P_{,X}=0$, so one 
further requires
 \beq \label{NoDegeEq}
  P_{,X} > 0.
 \eeq

Writing the principal part of eq. (\ref{CovFieldEq})
in terms of the ``effective'' inverse
metric
$ 
  G^{\mu\nu} = g^{\mu\nu}+2 {P_{,XX} \ov P_{,X}} {\nabla^\mu\phi\nabla^\nu\phi \ov M_f^4} ,
$ 
one can see that the equation is hyperbolic, and thus
defines a well-posed Cauchy problem \cite{Adams:2006sv, Rendall:2005fv},
if
 \beq \label{HypCau}
  P_{,X}+2 {\nabla^\mu\phi\nabla_\mu\phi \ov M_f^4} P_{,XX} > 0.
 \eeq

Moreover, the lightcone of the metric
$G_{\mu\nu}$
is inside or on the light cone of the metric
$g_{\mu\nu}$ if
 \beq \label{SubLum}
  {P_{,XX} \ov P_{,X}} \leq 0 \Rightarrow P_{,XX} \leq 0,
 \eeq
where at the end we have used (\ref{NoDegeEq}). If (\ref{SubLum}) is
satisfied no superluminal motion is possible.

These basic physical requirements
have nothing
to do with inflation in itself, but they will
lead to important consequences for it.
Let us stress that while the conditions from the
NEC and well-posed Cauchy problem are rather
basic, (\ref{SubLum})
seems not as fundamental. However, there is an obstruction to UV completion
if it is violated \cite{Adams:2006sv}, thus we adopt it.

Finally, a softly broken shift-symmetry 
also
constrains the dependence on
$\phi$ of $P(\phi, X)$ and $V(\phi)$. 
Indeed, from the almost conserved Noether
current, one finds that 
$|\partial_\phi V|,
M_f^4|\partial_\phi P| \propto \lambda_{sb}^n$, $n$ positive, with
$\lambda_{sb}$('s)$\ll 1$
the symmetry breaking parameter(s).

\subsection{Self-consistency of the theory}\label{ConsistTrunc}

In the effective theory truncated to
finite order in derivatives, the only trustworthy solutions are those
where higher derivatives are subdominant \cite{Burgess:2014lwa}. However, a generic
action as
 (\ref{GenActFD}) would typically admit runaway solutions
not satisfying this condition. This is
unsuitable to study onset and robustness of inflation,
as one would not be able to consider
generic initial conditions, according to the bounds (\ref{GenInCond}),
and trust all solutions.
Hence, one must make sure that higher-order derivatives
are subdominant for the relevant space of solutions, up to
the generic initial conditions (\ref{GenInCond}), close to the proper
cutoff $|X| \lesssim 1$ (but
within the validity limits of our
effective theory).
The consequences of this have typically been neglected in the literature
\cite{InitialConditionsIssue, InhomogeneousInitialConditionsIssue}.

Minimal conditions on the action that ensure this
behavior can be found by looking at the field equations (\ref{DcDynSis}).
Second derivatives can be ignored up to $|X| \lesssim 1$ provided
 \beq \label{TruncVal}
  c_s^2|_{|X| \lesssim 1} \ll 1 \,\, \Rightarrow\,\, -{P_{,XX} \ov P_{,X}}|_{|X| \lesssim 1} \gg 1.
 \eeq

Note that this is precisely what happens with the DBI action \cite{DBIInf}.
However, that case is peculiar since in some of the models inflation itself
occurs near the cutoff region, leading to strong constraints on the models.

We demand instead that $c_s^2 \!\ll \! 1$ {\em not} during
inflation, but 
around $|X| \! \lesssim \! 1$, to avoid
strong coupling of inflationary perturbations and large non
Gaussianities, which would generally invalidate the inflationary
model(s) \cite{ModifiedSoundSpeedBG}.

For the
system
(\ref{DcDynSis}), the condition
$c_s^2|_{|X| \lesssim 1} \sim 0$ implies
that the trajectories leave  the region
$|X| \!\lesssim\! 1$ {\em slowly}, since $\dot y \!\sim\! 0$.
As we shall see, this affects
inflation.

\subsection{Inflation
and
high-derivative corrections}\label{FinResCorrActi}

We list the most relevant results, in order of concern.
\paragraph{i) Attractors and fixed points.}
Because
of conditions (\ref{NoDegeEq}) and (\ref{HypCau}), the only fixed
points of the high-energy corrected system (\ref{DcDynSis}) are
    \beq \label{FixPoinHighDer}
      \{ (\phi, y) = (\phi_c, 0) \, | \, \partial_\phi \rho(\phi_c, 0) = 0\}.
    \eeq
These are equivalent to the fixed points of the standard system
(\ref{StanDynSis}) since
$\partial_\phi \rho|_{(\phi_c, 0)} \!\!=\!\! \partial_\phi V|_{\phi_c}$.
This is comforting, as it excludes the possibility for the solutions
to end up in different attractors and
goes in the direction
of supporting the standard dynamics.

New exact $y$-stationary points would
require $\partial_X P|_{(\phi, y)_{\text{new}}}\!\!=\!\!0$ as well as
$\partial_\phi \rho|_{(\phi, y)_{\text{new}}}\!\!=\!\!0$, but this is prevented by the
well-posedness of the field equations, see in particular eq. (\ref{NoDegeEq}).
Approximate $y$-stationary cases
($(\partial_X P, \partial_X \rho)|_{(\phi, y)_{\text{new}}}\!\! \ll\!\! 1$) are
possible, but would generally possess unsuitable features
(such as a sound speed
$c_s^2 \!\ll\! 1$ for the perturbation, which would have left potentially
observable imprints \cite{ModifiedSoundSpeedBG, ModifiedSoundSpeed}).

Approximate inflationary
solutions can be found with the method
described for the standard action.

The region $\mathcal{S}$ (where $\dot \phi$ is stationary) is now defined by
   \beq \label{ZeroLocHigDerAct}
      -\sqrt{3} c_s^2{\sqrt{2 P_{,X} y^2 + M_f^4 P + V} \ov M_{\text{Pl}}}y
     + {M_f^4 \ov 2}{\partial_\phi \rho(\phi) \ov \partial_X \rho(\phi)} = 0,
    \eeq
and the regions $\mathcal{D}/\mathcal{I}$ of
   decresing/increasing $\dot\phi$, occur where the left-hand side of
   (\ref{ZeroLocHigDerAct}) is negative/positive.

The curves
$y_{_\ast}(\!\phi\!) \!\in\! \mathcal{S}$ approximate actual solutions when
 \beq
  |y_\ast\partial_\phi y_\ast| \ll c_s^2\sqrt{3}{\sqrt{2 P_{,X} y_\ast^2 + M_f^4 P + V} \ov M_{\text{Pl}}}|y_\ast|,
 \eeq
   which replaces condition (\ref{SlRoIS}) of the standard scenario.
   These approximate solutions are inflationary when
   $2P_{,X} y^2+ M_F^4 P \ll V$.
As before, the global or local attractor nature of the inflationary solutions depends
on that of the fixed points (\ref{FixPoinHighDer}), which are also
part of $\mathcal{S}$.

\paragraph{ii) Onset of inflation.}
To start potential-driven inflation one typically
   needs lower values of $\dot\phi^2$
   than when using the standard action, since the kinetic
   energy density increases more rapidly with $\dot\phi^2$
   than the standard case
   because of conditions (\ref{HypCau}),
   (\ref{SubLum}), (\ref{TruncVal}),
while shift-symmetry makes
   it weakly dependent on $\phi$.
Hence,
solutions starting with generic initial conditions may reach the
inflationary regime too late.
Next we present
yet another effect that contributes even more
to this issue.

\paragraph{iii) Duration of inflation.}
The most relevant results concern the duration of inflation
and related predictions/implications.

The consistency conditions ensuring that
higher derivatives are
subdominant in the effective theory truncated at finite order in derivatives
imply that solutions flow more slowly out of the region
$|X| \lesssim 1$, see eq. (\ref{TruncVal}) and
comments.
Indeed, eqs.~(\ref{TruncVal}),
(\ref{SubLum}) and the weak dependence of $P$ on $\phi$
lead to an approximate solution for large $y$ and $V$ subdominant in
the energy density:
{\small \beq \label{ApproxSolCutX1}
 \text{sign(y)}\Bigl[P_{,_{X}}(-{y^2 \ov M_f^4})^{^{-{1 \ov 2}}}-P_{,_{X}}(-1)^{^{-{1 \ov 2}}}\Bigr]
    \! \propto \! {\sqrt{3} \ov M_{\text{Pl}}}(\phi-\phi_{\text{in}}).
 \eeq}
They also show that
{\small $P_{,_{X}}(\text{-}{y^2 \ov M_f^4})^{{-{1 \ov 2}}}\text{-}P_{,_{X}}(\text{-1})^{{-{1 \ov 2}}}\!\! >\!\! \text{-}\log({|y| \ov M_f^2})$}
(indeed,
using eq. (\ref{TruncVal}), one sees that typically the
right-hand side grows much more rapidly than the logarithm when $|y|$
decreases from $\lesssim\! M_f\!$).
Thus, comparing eq. (\ref{ApproxSolCutX1}) and the logarithm of (\ref{ExpoDamp}),
one can see that low values (inflationary regime) of $y=\dot \phi$
are typically attained more slowly than
the exponential damping found when higher-derivative corrections are not taken into account.
This slower motion would
extend beyond
the cutoff region, depending on
the explicit form of $P(\phi, X)$.

Due to this
slower damping of the kinetic energy,
the inflationary
attractor would be reached generally later than when
derivative corrections are not accounted for.
The duration of inflation could then be insufficient
to provide the desired effects on the subsequent universe evolution
\cite{Inflation, InflationReview},
or lead to observables signatures \cite{Cicoli:2014bja}.

\vspace{0.8cm}

The author thanks Anupam Mazumdar, and Gary Shiu for discussions,
and especially Augusto Sagnotti for discussions and for reading and
commenting this manuscript.
The author is supported by the Belgian National “Fond de la
Recherche Scientifique” F.R.S.-F.N.R.S. with a contract “charg\'e de recherche”.




\begin{thebibliography}{99}

\bibitem{Inflation}

  A.~H.~Guth,
  Phys.\ Rev.\ D  23 (1981) 347
  ;
   A.~D. Linde,
  Phys.Lett. B108 (1982) 389--393
  ;
   A.~Albrecht and P.~J. Steinhardt,
  Phys.Rev.Lett. 48 (1982) 1220--1223
  .

\bibitem{InflationReview}

  V.~Mukhanov,
  Cambridge, UK: Univ. Pr. (2005) 421 p
  ;
  S.~Weinberg,
  Oxford, UK: Oxford Univ. Pr. (2008) 593 p
  ;
  D.~H.~Lyth and A.~R.~Liddle,
  Cambridge, UK: Cambridge Univ. Pr. (2009) 497 p
  ;
  A.~D.~Linde,
  Contemp.\ Concepts Phys.\   5 (1990) 1
  [hep-th/0503203]
  ;
  D.~Baumann and L.~McAllister,
  arXiv:1404.2601 [hep-th]
  .
\bibitem{QuantumPerturbations}

  V.~F. Mukhanov and G.~V. Chibisov,
  JETP Lett. 33 (1981) 532--535
  ;
  G.~Chibisov and V.~F. Mukhanov,
  Mon.Not.Roy.Astron.Soc. 200 (1982) 535--550
  ;
  A.~H. Guth and S.~Pi,
  Phys.Rev.Lett. 49 (1982) 1110--1113
  ;
  S.~Hawking,
  Phys.Lett. B115 (1982) 295
  ;
  A.~A. Starobinsky,
  Phys.Lett. B117 (1982) 175--178
  ;
  J.~M. Bardeen, P.~J. Steinhardt, and M.~S. Turner,
  Phys.Rev. D28 (1983) 679
  .

\bibitem{InitialConditionsIssue}

  S.~Bird, H.~V.~Peiris and R.~Easther,
  Phys.\ Rev.\ D  78 (2008) 083518
  [arXiv:0807.3745 [astro-ph]]
  ;
  K.~Dutta, P.~M.~Vaudrevange and A.~Westphal,
  JCAP  1201 (2012) 026
  [arXiv:1109.5182 [hep-th]]
  ;
  S.~Downes, B.~Dutta and K.~Sinha,
  Phys.\ Rev.\ D  86 (2012) 103509
  [arXiv:1203.6892 [hep-th]]
  ;
  R.~Kallosh and A.~Linde,
  JCAP  1312 (2013) 006
  [arXiv:1309.2015 [hep-th]]
  ;
  R.~Kallosh, A.~Linde and D.~Roest,
  Phys.\ Rev.\ Lett.\   112 (2014) 1,  011303
  [arXiv:1310.3950 [hep-th]]
  ;
  A.~Corichi and D.~Sloan,
  Class.\ Quant.\ Grav.\   31 (2014) 062001
  [arXiv:1310.6399 [gr-qc]]
  ;
  V.~Faraoni,
  Phys.\ Lett.\ A  269 (2000) 209
  [gr-qc/0004007]
  ;
  D.~S.~Salopek and J.~R.~Bond,
  Phys.\ Rev.\ D  42 (1990) 3936
  ;
  L.~A.~Kofman, A.~D.~Linde and A.~A.~Starobinsky,
  Phys.\ Lett.\ B  157 (1985) 361
  ;
  V.~A.~Belinsky, I.~M.~Khalatnikov, L.~P.~Grishchuk and Y.~.B.~Zeldovich,
  Phys.\ Lett.\ B  155 (1985) 232
  ;
  A.~Ijjas, P.~J. Steinhardt, and A.~Loeb,
  Phys.Lett. B723 (2013) 261--266,
  arXiv:1304.2785
  ;
  A.~H. Guth, D.~I. Kaiser, and Y.~Nomura,
  Phys.Lett. B733 (2014) 112--119,
  arXiv:1312.7619
  ;
   A.~Linde,
   arXiv:1402.0526
  ;
  A.~Ijjas, P.~J. Steinhardt, and A.~Loeb,
  arXiv:1402.6980
  .


\bibitem{Remmen:2013eja}
  G.~N.~Remmen and S.~M.~Carroll,
  Phys.\ Rev.\ D  88 (2013) 083518
  [arXiv:1309.2611 [gr-qc]]
  .

\bibitem{InhomogeneousInitialConditionsIssue}

  J.~H.~Kung and R.~H.~Brandenberger,
  Phys.\ Rev.\ D  40 (1989) 2532
  ;
  S.~R. Coleman and E.~J. Weinberg,
  Phys.Rev. D7 (1973) 1888--1910
  ;
  D.~S. Goldwirth,
  Phys.Rev. D43 (1991) 3204--3213
  ;
  D.~S. Goldwirth and T.~Piran,
  Phys.Rept. 214 (1992) 223--291
  ;
  A.~Albrecht, R.~H. Brandenberger, and R.~Matzner,
  Phys.Rev. D32 (1985) 1280
  ;
  J.~Kung and R.~H. Brandenberger,
  Phys.Rev. D40 (1989) 2532
  ;
  R.~H. Brandenberger and J.~Kung,
  Phys.Rev. D42 (1990) 1008--1015
  ;
  H.~A. Feldman and R.~H. Brandenberger,
  Phys.Lett. B227 (1989) 359
  ;
  R.~Easther, L.~C.~Price and J.~Rasero,
  arXiv:1406.2869 [astro-ph.CO]
  .

\bibitem{ShiftSymmetry}

  F.~C. Adams, J.~R. Bond, K.~Freese, J.~A. Frieman, and A.~V. Olinto,
  Phys.Rev. {\bf D47} (1993) 426--455,
  [arXiv:hep-ph/9207245 [hep-ph]]
  ;
  N.~Arkani-Hamed, H.~-C.~Cheng, P.~Creminelli and L.~Randall,
  JCAP  0307 (2003) 003
  [hep-th/0302034]
  ;
  P.~Brax and J.~Martin,
  Phys.\ Rev.\ D  72 (2005) 023518
  [hep-th/0504168]
  .


\bibitem{ArmendarizPicon:1999rj}
  C.~Armendariz-Picon, T.~Damour and V.~F.~Mukhanov,
  Phys.\ Lett.\ B {\bf 458} (1999) 209
  [hep-th/9904075]
  .

\bibitem{GhostInflation}

  L.~Senatore,
  Phys.\ Rev.\ D {\bf 71} (2005) 043512
  [astro-ph/0406187]
  ;
  N.~Arkani-Hamed, P.~Creminelli, S.~Mukohyama and M.~Zaldarriaga,
  JCAP {\bf 0404} (2004) 001
  [hep-th/0312100]
  .


\bibitem{Adams:2006sv}
  A.~Adams, N.~Arkani-Hamed, S.~Dubovsky, A.~Nicolis and R.~Rattazzi,
  JHEP  0610 (2006) 014
  [hep-th/0602178]
  .

\bibitem{Supersymmetry}

  E.~J. Copeland, A.~R. Liddle, D.~H. Lyth, E.~D. Stewart, and D.~Wands,
  Phys.Rev. D49 (1994) 6410--6433,
  arXiv:astro-ph/9401011 
  ;
  D.~Baumann and D.~Green,
  Phys.Rev. D85 (2012) 103520,
  arXiv:1109.0292 [hep-th]
  .

\bibitem{RemediesShiftSymmetry}

  N.~Kaloper and L.~Sorbo,
  Phys.\ Rev.\ Lett.\  {\bf 102} (2009) 121301
  [arXiv:0811.1989 [hep-th]]
  ;
  N.~Kaloper and A.~Lawrence,
  arXiv:1404.2912 [hep-th]
  ;
  S.~Dimopoulos, S.~Kachru, J.~McGreevy and J.~G.~Wacker,
  JCAP {\bf 0808}, 003 (2008)
  [hep-th/0507205]
  ;
  L.~McAllister, E.~Silverstein and A.~Westphal,
  Phys.\ Rev.\ D {\bf 82} (2010) 046003
  [arXiv:0808.0706 [hep-th]]
  ;
  F.~Marchesano, G.~Shiu and A.~M.~Uranga,
  arXiv:1404.3040 [hep-th]
  .

\bibitem{Chialva:2014rla}
  D.~Chialva and A.~Mazumdar,
  arXiv:1405.0513 [hep-th]
  .

\bibitem{PerturbationCorrections}

  D.~Chialva,
  PoS EPS-HEP2013 (2013) 478
  [arXiv:1312.2899 [hep-th]]
  ;
  D.~Chialva,
  JCAP  1210 (2012) 037
  [arXiv:1108.4203 [astro-ph.CO]]
  ;
  D.~Chialva,
  JCAP  1201 (2012) 037
  [arXiv:1106.0040 [hep-th]]
  .

\bibitem{ModifiedSoundSpeedBG}

  D.~Baumann and D.~Green,
  JCAP  1109 (2011) 014
  [arXiv:1102.5343 [hep-th]]
  .

\bibitem{ModifiedSoundSpeed}

  J.~Khoury and F.~Piazza,
  JCAP  0907 (2009) 026
  [arXiv:0811.3633 [hep-th]]
  ;
  M.~Nakashima, R.~Saito, Y.~-i.~Takamizu and J.~'i.~Yokoyama,
  Prog.\ Theor.\ Phys.\   125 (2011) 1035
  [arXiv:1009.4394 [astro-ph.CO]]
  ;
  M.~Park and L.~Sorbo,
  Phys.\ Rev.\ D  85 (2012) 083520
  [arXiv:1201.2903 [astro-ph.CO]]
  ;
  A.~Achucarro, J.~-O.~Gong, S.~Hardeman, G.~A.~Palma and S.~P.~Patil,
  JHEP  1205 (2012) 066
  [arXiv:1201.6342 [hep-th]]
  ;
  A.~Achucarro, V.~Atal, S.~Cespedes, J.~-O.~Gong, G.~A.~Palma and S.~P.~Patil,
  Phys.\ Rev.\ D  86 (2012) 121301
  [arXiv:1205.0710 [hep-th]]
  ;
  A.~Achúcarro, J.~-O.~Gong, G.~A.~Palma and S.~P.~Patil,
  Phys.\ Rev.\ D  87 (2013) 12,  121301
  [arXiv:1211.5619 [astro-ph.CO]]
  ;
  A.~Ashoorioon, D.~Chialva, U.~Danielsson,
  JCAP {\bf 1106 } (2011)  034.
  [arXiv:1104.2338 [hep-th]].

\bibitem{Overshooting}

  R.~Brustein and P.~J.~Steinhardt,
  Phys.\ Lett.\ B  302 (1993) 196
  [hep-th/9212049]
  ;
  B.~Underwood,
  Phys.\ Rev.\ D  78 (2008) 023509
  [arXiv:0802.2117 [hep-th]]
  ;
  S.~Bird, H.~V.~Peiris and D.~Baumann,
  Phys.\ Rev.\ D  80 (2009) 023534
  [arXiv:0905.2412 [hep-th]]
  .

\bibitem{EnergyConditionInstability}

  R.~V.~Buniy, S.~D.~H.~Hsu and B.~M.~Murray,
  Phys.\ Rev.\ D  74 (2006) 063518
  [hep-th/0606091]
  ;
  I.~Y.~.Aref'eva and I.~V.~Volovich,
  Theor.\ Math.\ Phys.\   155 (2008) 503
  [hep-th/0612098]
  ;
  V.~A.~Rubakov,
  Phys.\ Usp.\   57 (2014) 128
  [arXiv:1401.4024 [hep-th]]
  ;
  E.~Babichev, V.~Mukhanov and A.~Vikman,
  JHEP  0802 (2008) 101
  [arXiv:0708.0561 [hep-th]]
  .

\bibitem{NECViolation}

  B.~Elder, A.~Joyce and J.~Khoury,
  Phys.\ Rev.\ D {\bf 89} (2014) 044027
  [arXiv:1311.5889 [hep-th]]
  ;
  V.~A.~Rubakov,
  Phys.\ Usp.\  {\bf 57} (2014) 128
  [arXiv:1401.4024 [hep-th]]
  .


\bibitem{Rendall:2005fv}
  A.~D.~Rendall,
  Class.\ Quant.\ Grav.\   23 (2006) 1557
  [gr-qc/0511158]
  .

\bibitem{Burgess:2014lwa}
  C.~P.~Burgess and M.~Williams,
  JHEP {\bf 1408} (2014) 074
  [arXiv:1404.2236 [gr-qc]].

\bibitem{DBIInf}

  E.~Silverstein and D.~Tong,
  Phys.Rev. D70 (2004) 103505,
  arXiv:hep-th/0310221 [hep-th]
  ;
  M.~Alishahiha, E.~Silverstein, and D.~Tong,
  Phys.Rev. D70 (2004) 123505,
  arXiv:hep-th/0404084 [hep-th]
  .



\bibitem{Cicoli:2014bja}
  M.~Cicoli, S.~Downes, B.~Dutta, F.~G.~Pedro and A.~Westphal,
  arXiv:1407.1048 [hep-th]
  .

\end{thebibliography}
\end{document}